\documentclass[aps,pra,showpacs]{revtex4}

\usepackage{graphicx}
\begin{document}
\title{Uncertainty limits for quantum metrology obtained from\\ the statistics of weak measurements}

\author{Holger F. Hofmann}
\email{hofmann@hiroshima-u.ac.jp}
\affiliation{
Graduate School of Advanced Sciences of Matter, Hiroshima University,
Kagamiyama 1-3-1, Higashi Hiroshima 739-8530, Japan}
\affiliation{JST,Crest, Sanbancho 5, Chiyoda-ku, Tokyo 102-0075, Japan
}

\begin{abstract}
Quantum metrology uses small changes in the output probabilities of a quantum measurement to estimate the magnitude of a weak interaction with the system. The sensitivity of this procedure depends on the relation between the input state, the measurement results, and the generator observable describing the effect of the weak interaction on the system. This is similar to the situation in weak measurements, where the weak value of an observable exhibits a symmetric dependence on initial and final conditions. In this paper, it is shown that the phase sensitivity of a quantum measurement is in fact given by the variance of the imaginary parts of the weak values of the generator over the different measurement outcomes. It is then possible to include the limitations of a specific quantum measurement in the uncertainty bound for phase estimates by subtracting the variance of the real parts of the weak values from the initial generator uncertainty. This uncertainty relation can be interpreted as the time-symmetric formulation of the uncertainty limit of quantum metrology, where the real parts of the weak values represent the information about the generator observable in the final measurement result. 
\end{abstract}


\maketitle

\section{Introduction}
Quantum metrology is the application of quantum measurements to the problem of determining a classical parameter that describes the magnitude of a well-defined action on the quantum system. Since this problem avoids many of the ambiguities associated with quantum measurements and instead presents a clear practical objective, the field is attracting a lot of attention, especially from quantum optics where experimental applications to lithography and precision measurements such as gravitational wave detection are within the reach of present technologies \cite{Gio04,Res07,Nag07,Hof07,Pez08,Oka08,Sun08,Dem09,Dur10,Ani10,Afe10}. Nevertheless all practical applications of quantum systems also reflect the underlying statistical structure that distinguishes quantum physics from classical physics. In quantum metrology, this relation is most clearly expressed by the uncertainty limit of phase estimation that identifies the maximal phase sensitivity of a quantum state with the uncertainty of the generator of the unitary operation representing the phase shift \cite{Bra94,Gio06,Hof09}. Such a relation between uncertainty and precision would be unthinkable in classical physics, where optimal precision is naturally obtained when all uncertainties vanish. 

At first sight, the uncertainty principle appears to represent the perfect black box, seemingly frustrating any attempt to uncover details of the underlying physics. However, a unique way of beating the uncertainty principle was discovered in 1988 by Aharonov, Albert and Vaidman \cite{Aha88}: statistical information about a system can even be obtained if the measurement interaction is so weak that the back-action on the system can be neglected. The average value of the weak measurement then reveals a time-symmetric dependence on initial and final conditions, indicating that a more complete description of quantum statistics is possible \cite{Aha01,Shi10,Hof10}. However, the weak values obtained by post-selecting a final measurement outcome can lie far outside the range of possible eigenvalues \cite{Aha88}. As a result, the interpretation of weak values caused some controversy \cite{Duc89,Leg89}. Still, the practical validity of the experimental prediction was never in question, and there have been numerous experimental demonstrations of the effect \cite{Rit91,Pry05,Mir07,Lun09,Yok09}. In particular, it has been shown that weak values exceeding the maximal eigenvalue of the observable can be used to amplify the effect of an interaction involving this observable, thus making it easier to measure small effects \cite{Hos08,Dix09,Sta09}. This amplification effect by itself already suggests some relation between weak values and the sensitivity of parameter estimation. In a recent study, weak measurements were therefore compared with interferometry, resulting in the suggestion that the imaginary part of the weak value may be useful for the enhancement of sensitivity \cite{Bru10}. However, surprisingly little attention has been paid to the more fundamental relation between parameter estimation and the weak values of generator observables.

In the following, it is pointed out that the situation studied in quantum metrology is in fact identical to the measurement procedure used to define the imaginary part of the weak value \cite{Jos07}. Therefore, the phase estimation problem can be formulated in terms of the complex weak values of the generator observable obtained for the different outcomes of the final measurement by which the phase estimate is obtained. In this formulation, the phase sensitivity of a final measurement is given by the fluctuation of the imaginary parts of the weak values. Since the total fluctuation of the complex weak values cannot exceed the uncertainty of the initial state, it is easy to see that phase estimation must be limited by the generator uncertainty. Moreover, any fluctuations of the real part of the weak values necessarily reduces the phase sensitivity that can be achieved in the quantum measurement by the corresponding amount of uncertainty. As will be discussed in more detail below, the fluctuations of the real part of the weak values can be interpreted as a reduction in the initial uncertainty of the generator observable when the final measurement result is used to estimate the value of the observable instead of the parameter. It is therefore possible to formulate a time-symmetric uncertainty limit of phase estimation that treats the information gained about the generator observable in the final measurement on the same level as the information initially available in the input state. The maximal phase sensitivity is then equal to the uncertainty about the generator observable that remains {\it after} the final measurement. The time-symmetric uncertainty limit of phase estimation thus indicates a fundamental complementarity between the dynamics represented by the phase parameter of a unitary evolution and the accessible information about the generator observable of that unitary.

\section{Imaginary weak values as logarithmic derivatives of phase dependent probabilities}
In the typical quantum metrology setup, a quantum state $\mid \psi \rangle$ is modified by a unitary transformation $\hat{U}=\exp(-i \phi \hat{A})$ defined by the self-adjoint generator $\hat{A}$ and an unknown phase shift $\phi$. The goal is to estimate the phase shift from the measurement statistics observed in the output using a quantum measurement. Since the limit of sensitivity is given by the ability to detect small phase differences, it is usually sufficient to consider the case of small phase shifts $\phi$ around a phase value defined as $\phi=0$. The phase sensitivity is then given by the differential dependence of the measurement probabilities $p(m)$ on the phase shift $\phi$. In general, phase sensitivities at other phase values can be obtained by rotating the quantum state $\mid \psi \rangle$ to the corresponding initial phase and analysing the differential effects of small shifts around this phase value.

In the following, I will consider the phase sensitivity achieved by a standard von Neumann measurement represented by projections on an orthogonal basis set $\{ \mid m \rangle \}$. For a specific measurement, the problem is essentially a classical one, defined by the dependence of the probabilities $p(m)=|\langle m \mid \hat{U} \mid \psi \rangle|^2$ on the parameter $\phi$. As is well known, the optimal phase estimation strategy for this classical statistical problem achieves a phase sensitivity given by the Cramer-Rao bound. If the phase sensitivity is expressed in terms of the inverse quadratic error of the phase estimate for a specific quantum measurement, the Cramer-Rao bound is equal to the Fisher information of the phase dependent probabilities, as given by the right hand side of the following equality
\begin{equation}
\label{eq:Fisher}
\frac{1}{\delta \phi_{\mathrm{min.}}^2}=\sum_m \left(\frac{\partial \ln(p(m))}{\partial \phi} \right)^2  p(m).
\end{equation}
It is possible to interpret this expression of phase sensitivity as a statistical variance of an $m$-dependent estimator defined by the logarithmic derivative of $p(m)$ in $\phi$. We can now use the quantum mechanical definition of the probability $p(m)$ in terms of the initial state $\mid \psi \rangle$, the final state $\mid m \rangle$, and the generator of the unitary $\hat{A}$ to determine the $m$-dependent value of this logarithmic derivative. The result reads 
\begin{equation}
\label{eq:logderiv}
\frac{\partial \ln(p(m))}{\partial \phi} = 2 \mathrm{Im}
\left(
\frac{\langle m \mid \hat{A} \mid \psi \rangle}{\langle m \mid \psi \rangle}
 \right).
\end{equation}
Due to the time reversal symmetry of unitary transformations, this quantum mechanical expression for the logarithmic derivative of the output probability is anti-symmetric with regard to an exchange of initial state $\mid \psi \rangle$ and final state $\mid m \rangle$. In this sense, the logarithmic derivatives can be interpreted as time-symmetric contributions of the trajectory jointly defined by initial conditions $\mid \psi \rangle$ and final conditions $\mid m \rangle$ to the phase estimation procedure. 

In fact, the logarithmic derivative of the output probability is equal to the imaginary part of the weak value of the generator $\hat{A}$ observed for an initial state $\mid \psi \rangle$ and a final state $\mid m \rangle$ \cite{Aha88,Jos07}. To understand the reason for this fundamental relation between phase sensitivity and imaginary weak values, it is useful to take a closer look at the definition of weak values. Experimentally, the weak values of an operator $\hat{A}$ can be obtained by observing the average shift of a meter system induced by a weak interaction generated by the Hamiltonian given by $g \hat{p}\hat{A}$, where $g$ is a coupling constant and $\hat{p}$ is the momentum of the meter system. If no final measurement is performed, the average of the weak measurement results corresponds to the expectation value of $\hat{A}$ in the initial state $\mid \psi \rangle$. However, it is possible to define the system more precisely by performing a final measurement projecting the state onto the basis states $\{ \mid m \rangle \}$. The average result of the weak measurement conditioned by the final measurement result $m$ is then equal to the real part of the weak value,
\begin{equation}
\label{eq:weakvalue}
\langle \hat{A} \rangle_{\mathrm{weak}}(m) = \frac{\langle m \mid \hat{A} \mid \psi \rangle}{\langle m \mid \psi \rangle}.
\end{equation}
Since the meter shift is a real number, the imaginary part of the weak value does not appear in the average meter position $\hat{x}$ measured after the weak system-meter interaction. However, it has been pointed out that the imaginary part of the weak value can be observed in a measurement of the meter momentum $\hat{p}$ \cite{Jos07}. In the formal mathematical description, this ``shift'' in momentum looks just like the shift in pointer position caused by the real weak values. However, a momentum measurement commutes with the interaction Hamiltonian, indicating that the measurement can be interpreted as a selection of an unshifted momentum component from the statistics of the initial meter state. The momentum eigenvalue $p$ observed in the meter measurement can then be treated like a classical random variable that determines the phase of a unitary operation acting only on the system. For an interaction time of $\delta t$, the phase of this unitary is $\phi = g \,p \, \delta t$. The imaginary weak value is then obtained by determining the correlations between the momentum eigenvalue $p$ and the change in probability for a specific outcome $m$. Effectively, the measurement of imaginary weak values proposed in \cite{Jos07} is therefore a direct experimental determination of the logarithmic derivative of the output probability $p(m)$ using randomly varying phase shifts. 

\section{Variance of complex weak values and time-symmetric uncertainty} 
Eqs.(\ref{eq:Fisher}) and (\ref{eq:logderiv}) show that the phase sensitivity of a quantum measurement corresponds to the fluctuations of the imaginary weak values for different measurement outcomes $m$. In general, each measurement outcome $m$ is associated with its own complex weak value, as given by eq.(\ref{eq:weakvalue}). In this sense, a measurement of $m$ is also a measurement of $\hat{A}$, even if the states $\mid m \rangle$ are not eigenstates of $\hat{A}$.  
An important observation in this context is due to Hosoya and Shikano \cite{Hosoya10}: regardless of the measurement performed, the variance of the complex weak values of any observable is exactly equal to the uncertainty of the same observable in the initial pure state,
\begin{equation}
\label{eq:quasideterm}
\sum_m \left|\langle \hat{A} \rangle_{\mathrm{weak}}(m)-\langle \hat{A} \rangle \right|^2 p(m) = \Delta A_{\mathrm{in.}}^2,
\end{equation}
where the average of the weak values obtained for different measurement outcomes $m$ is always equal to the expectation value $\langle \hat{A} \rangle$ in the initial state $\mid \psi \rangle$. This means that the distribution of complex weak values obtained for any final measurement $\{ \mid m \rangle\}$ has the same variance as the distribution of eigenvalues obtained when $\mid m \rangle$ are eigenstates of $\hat{A}$, even though the individual weak values do not correspond to the eigenvalues of $\hat{A}$. 

With regard to phase estimation, the variance of the complex weak values can be separated into a contribution from the real parts of the weak values and a contribution from the imaginary part of the weak values. According to Eqs.(\ref{eq:Fisher}) and (\ref{eq:logderiv}), the variance of the imaginary weak values is equal to the phase sensitivity of the quantum measurement. For pure states, the phase sensitivity is therefore given by 
\begin{equation}
\label{eq:purebound}
\frac{1}{\delta \phi^2} \; = \; 4 \Delta A_{\mathrm{in.}}^2 -
4 \sum_m \left(\mathrm{Re}\left[\langle \hat{A} \rangle_{\mathrm{weak}}(m)\right]-\langle \hat{A} \rangle \right)^2 p(m).
\end{equation}
This equality indicates that, in the pure state limit, quantum measurements are optimal for phase estimation when the real weak values of the generator are all equal to the initial expectation value \cite{phase}. 

Measurement theory provides an intuitive interpretation of the real weak values as the best possible estimate of $\hat{A}$ based on the final measurement result $m$ \cite{Oza03,Hal04,Joh04}. Classical statistics then suggests that the average uncertainty of the estimate of $\hat{A}$ is given by the difference between the initial uncertainty and the fluctuations of the estimate. In terms of the quantum formalism, this final uncertainty can be defined as \cite{Oza03}
\begin{eqnarray} 
\label{eq:finalDA}
\Delta A_{\mathrm{est.}}^2 &=& \sum_m \langle \psi \mid \left(\hat{A}-\mathrm{Re}\left[\langle \hat{A} \rangle_{\mathrm{weak}}(m)\right]\right)^2 \mid \psi \rangle \; p(m)
\nonumber \\
&=& \Delta A_{\mathrm{in.}}^2 - \sum_m \left(\mathrm{Re}\left[\langle \hat{A} \rangle_{\mathrm{weak}}(m)\right]-\langle \hat{A} \rangle \right)^2 p(m).
\end{eqnarray}
Therefore, Eq.(\ref{eq:purebound}) defines the sensitivity of phase estimation for pure states as four times the generator uncertainty $\Delta A_{\mathrm{est.}}^2$ for estimates based on the final measurement outcome $m$. 

Significantly, the final uncertainty $ \Delta A_{\mathrm{est.}}^2$ represents the effects of the final measurement in terms of the information gain about $\hat{A}$. In particular, a precise measurement of $\hat{A}$ reduces the phase sensitivity to zero, while maximal phase sensitivity is achieved by a measurement that is completely insensitive to $\hat{A}$. Ultimately, phase sensitivity requires that both preparation and measurement are equally insensitive to the generator observable $\hat{A}$. It is therefore possible to interpret the uncertainty limit given by $\Delta A_{\mathrm{est.}}^2$ as the time-symmetric generalization of the initial state limit given by $\Delta A_{\mathrm{in.}}^2$.

\section{Time-symmetric uncertainty relation}

Eq.(\ref{eq:purebound}) shows that phase estimation and information about the generator observable are complementary for all pure states, such that the phase sensitivity is a deterministic function of the generator uncertainty in estimates based on the final measurement outcome. In a realistic situation, both the phase sensitivity and the information available about the generator $\hat{A}$ will be reduced by experimental noise. Intuitively, it is therefore clear that decoherence effects should reduce the sensitivity given by $1/\delta \phi^2$, while the final uncertainty given by $\Delta A_{\mathrm{est.}}^2$ should be increased. For the general case of mixed states and noisy measurements, the relation between phase sensitivity and final uncertainty of the generator $\hat{A}$ is therefore given by
\begin{equation}
\label{eq:newbound}
\frac{1}{\delta \phi^2} \leq 4 \Delta A_{\mathrm{est.}}^2,
\end{equation}
where the phase sensitivity $1/\delta \phi^2$ and the final uncertainty $\Delta A_{\mathrm{est.}}$ are both defined in terms of the experimentally accessible measurement statistics of the initial mixed state $\hat{\rho}$ and the positive operator valued measure (POVM) $\{\hat{\Pi}_m\}$ of the final measurement. Specifically, both can be expressed in terms of the complex weak values of the generator $\hat{A}$, given by
\begin{equation}
\label{eq:mixwv}
\langle \hat{A} \rangle_{\mathrm{weak}}(m) = \frac{\mbox{Tr}\left\{\hat{\Pi}_m \hat{A}\; \hat{\rho}\right\}}{\mbox{Tr}\left\{\hat{\Pi}_m \hat{\rho}\right\}}.
\end{equation}
As in the pure state case, the phase sensitivity is equal to four times the fluctuations of the imaginary weak values,
\begin{equation}
\frac{1}{\delta \phi^2} = 4 \sum_m \mathrm{Im}\left[\langle \hat{A} \rangle_{\mathrm{weak}}(m)\right]^2 \mbox{Tr}\left\{\hat{\Pi}_m \hat{\rho}\right\},
\end{equation}
and the final uncertainty of $\hat{A}$ can be given as
\begin{equation}
\Delta A_{\mathrm{est.}} = \mbox{Tr}\left\{\hat{A}^2\; \hat{\rho}\right\} - \sum_m \mathrm{Re}\left[\langle \hat{A} \rangle_{\mathrm{weak}}(m)\right]^2 \mbox{Tr}\left\{\hat{\Pi}_m \hat{\rho}\right\}.
\end{equation}

For the mathematical proof that the uncertainty relation (\ref{eq:newbound}) holds in general, it is sufficient to show that the average absolute square of the weak values cannot exceed the average of $\hat{A}^2$ in the initial state. This can be done by formulating the inequality as a Cauchy-Schwarz inequality for the vectors $\hat{\Pi}_m^{1/2} \hat{A} \hat{\rho}^{1/2}$ and $\hat{\rho}^{1/2}\hat{\Pi}_m^{1/2}$. The average over the squared complex weak values is then limited by
\begin{equation}
\label{eq:CS}
\sum_m \frac{\left|\mbox{Tr}\left\{\hat{\Pi}_m \hat{A}\; \hat{\rho}\right\} \right|^2}{\mbox{Tr}\left\{\hat{\Pi}_m \hat{\rho}\right\}} \leq \sum_m  \mbox{Tr}\left\{\hat{\Pi}_m \hat{A}\; \hat{\rho}\; \hat{A}\right\} = \mbox{Tr}\left\{\hat{A}^2 \; \hat{\rho}\right\}.
\end{equation}
Thus, noise in either the state preparation or the measurement will prevent the exact achievement of the uncertainty limit given by Eq.(\ref{eq:newbound}). 

A particularly interesting aspect of the present analysis is that it applies to noisy measurements. In a realistic phase estimation setup, it is equally challenging to realize a precise quantum measurement as it is to realize quantum state preparation. Therefore, the phase sensitivity given by the Fisher information of a quantum state obtained by optimizing the measurement is not necessarily achievable in realistic experiments. Eq.(\ref{eq:CS}) shows that the statistical operators representing measurement outcomes and initial states contribute symmetrically to the reduction of phase sensitivity below the uncertainty limit. The analysis of weak values for noisy measurements may therefore be helpful in evaluating general requirements for quantum measurements in terms of the achievable phase sensitivities, including a possible optimization of the input states based on the available measurement precision.
 
\section{Complementarity of dynamics and generator values}

On the fundamental level, the time-symmetric uncertainty limit of phase estimation differs from the conventional limit for optimized measurements because it includes the effects of measurement information about the generator in the final measurement. The final uncertainty $\Delta A_{\mathrm{est.}}^2$ therefore defines an uncertainty about the generator $\hat{A}$ that cannot be resolved by any future measurements. It is in effect the uncertainty of the whole history of the quantum system, between preparation and measurement. If the history of the quantum system is represented by an initial state $\mid \psi \rangle$ and a final measurement $\{ \mid m \rangle\}$, Eq.(\ref{eq:purebound}) shows that the initial uncertainty of $\hat{A}$ in $\mid \psi \rangle$ is completely resolved by the measurement, either as generator information given by the real weak values of $\hat{A}$, or as phase information given by the imaginary values of $\hat{A}$. Eq.(\ref{eq:purebound}) thus indicates a fundamental complementarity between the information obtained about the dynamics generated by $\hat{A}$ and the information about the real value of $\hat{A}$ in a specific measurement $\{ \mid m \rangle\}$, where the balance between the two kinds of information is given by the complex weak values. 

Ultimately, the measurement of the initial state is maximally sensitive to phase shifts generated by operators with real weak values equal to the initial expectation value, and completely insensitive to phase shifts generated by operators whose real weak values reproduce the fluctuations of a projective measurement of the eigenstates. For a given combination of initial pure state $\mid \psi \rangle$ and final measurement $\{ \mid m \rangle\}$, each operator $\hat{A}$ can be expanded in terms of the phase-adjusted final states 
\begin{equation}
\mid \gamma (m) \rangle = \frac{\langle m \mid \psi \rangle}{|\langle m \mid \psi \rangle|} \mid m \rangle. 
\end{equation} 
In this expansion, the complex weak value is given by
\begin{equation}
\label{eq:wvphase}
\langle \hat{A} \rangle_{\mathrm{weak}}(m) = 
\sum_{m^\prime} \langle \gamma (m) \mid \hat{A} \mid \gamma (m^\prime) \rangle
\left| \frac{\langle m^\prime \mid \psi \rangle}{\langle m \mid \psi \rangle}\right|.
\end{equation}
Since the phase factors originating from the overlaps between $\mid \psi \rangle$ and $\{ \mid m \rangle\}$ have been incorporated in the basis $\{ \mid \gamma(m) \rangle\}$, the real part of the weak value originates from the real parts of the matrix elements of $\hat{A}$, while the imaginary part originates from the imaginary parts. Since $\hat{A}$ is self-adjoint, the real part is given by the symmetric component of the matrix describing $\hat{A}$, while the imaginary part is given by the anti-symmetric components. It is therefore possible to separate the operator components describing dynamical changes in the trajectory from $\mid \psi \rangle$ to $\{ \mid m \rangle \}$ from the components describing physical properties determined by $\{ \mid m \rangle \}$. In particular, the phase sensitivity and the phase estimation procedure are completely independent of the real symmetric part of the generator $\hat{A}$, and the final measurement $\{ \mid m \rangle \}$ is insensitive to the phase shifts generated by any of its symmetric operators. Each combination of input state and measurement thus defines a separation of the $d^2$-dimensional operator space into $d(d-1)/2$ anti-symmetric generators of phase shifts observed in the measurement and $d(d+1)/2$ symmetric observables determines by the measurement. 

\section{Conclusions}

In conclusion, it has been shown that quantum metrology is fundamentally related to weak values, since the logarithmic derivative of the measurement probability $p(m)$ that plays a pivotal role in phase estimation is equal to twice the imaginary part of the weak values of the generator observable. As a result, the phase sensitivity of pure states is given by the fluctuations of the imaginary part of the weak values. Since the total fluctuations of the weak value are always equal to the inital generator uncertainty for pure states (and smaller for mixed states), it is possible to characterize the efficiency of a quantum measurement in terms of the division of weak value fluctuations into real and imaginary parts. The result is an uncertainty limit for phase estimation that includes the information about the generator observable obtained in the final measurement. 

The time-symmetric uncertainty limit of phase estimation differs from the conventional limit for the input states by including the statistical properties of the final quantum measurement. It is therefore not a statement about quantum states, but a statement about the complete phase estimation process from state preparation to measurement. The uncertainty then refers to a lack of information about the generator observable in the data classically available after all measurements have been completed. From the practical side, this may make it easier to describe the effects of measurement errors in the final measurement and to evaluate the experimental limitations of quantum measurements. On the fundamental side, it is interesting to note that each set of state preparation and measurement separates the space of operator observables into a set of generators and a set of observables, such that the phase estimate is optimal for the generators and impossible for the observables, while estimates of the observable itself are optimal for the observables and impossible for the generators. 
 
Ultimately, the connection between phase sensitivity and weak values could provide an essential link between the almost classical features of continuous dynamics and the less intuitive non-classical aspects of quantum statistics. The time-symmetric formulation of the uncertainty bound of quantum metrology may thus be helpful in linking and unifying different approaches in quantum information and quantum physics.

\section*{Acknowledgment}
Part of this work has been supported by the Grant-in-Aid program of the Japanese Society for the Promotion of Science, JSPS.


\begin{thebibliography}{xyz00}
\bibitem{phase}
It may be worth noting that this reduction of sensitivity due to non-optimal measurements was also discussed in \cite{Dur10}, where the correponding term is described as a variance of (quantum) phase velocities.
\end{thebibliography}
\end{document}